\documentclass[12pt,preprint]{aastex}
\usepackage{graphics}

\slugcomment{}
\shorttitle{X-ray Bo\"otes Survey}
\shortauthors{Murray et al.}

\begin{document}

\title{XBo\"otes: An X-ray Survey of the\\
NDWFS Bo\"otes Field--\\
Paper I -- Overview and Initial Results}

\author{Stephen S. Murray\altaffilmark{1}, Almus Kenter\altaffilmark{1},
William R. Forman\altaffilmark{1}, Christine Jones\altaffilmark{1},
\\
Paul J. Green\altaffilmark{1}, Christopher S. Kochanek\altaffilmark{5},
Alexey Vikhlinin\altaffilmark{1}, Daniel Fabricant\altaffilmark{1},
\\
 Giovani Fazio\altaffilmark{1}, Kate Brand\altaffilmark{2}, Michael
J. I. Brown\altaffilmark{2}\altaffilmark{,6}, Arjun Dey\altaffilmark{2},
Buell T. Jannuzi\altaffilmark{2}, \\
Joan Najita\altaffilmark{2}, Brian McNamara\altaffilmark{3}, Joseph
Shields\altaffilmark{3}, \\
and Marcia Rieke\altaffilmark{4}}

\begin{abstract}
We obtained a 5 ksec deep Chandra X-ray Observatory ACIS-I map of
the 9.3 square degree Bo\"otes field of the NOAO Deep Wide-Field
Survey. Here we describe the data acquisition and analysis strategies
leading to a catalog of 4642 (3293) point sources with 2 or more (4
or more) counts, corresponding to a limiting flux of roughly $4(8)\times10^{-15}\,\mathrm{erg\, cm^{-2}s^{-1}\,}$in
the 0.5-7 keV band. These Chandra XBo\"otes data are unique in that
they consitute the widest contiguous X-ray field yet observed to such
a faint flux limit. Because of the extraordinarily low background
of the ACIS, we expect only 14\% (0.7\%) of the sources to be spurious.
We also detected 43 extended sources in this survey. The distribution
of the point sources among the 126 pointings (ACIS-I has a 16 x 16
arcminute field of view) is consistent with Poisson fluctuations about
the mean of 36.8 sources per pointing. While a smoothed image of the
point source distribution is clumpy, there is no statistically significant
evidence of large scale filamentary structure. We do find however,
that for $\theta>1$ arcminute, the angular correlation function of
these sources is consistent with previous measurements, following
a power law in angle with slope $\sim-0.7$. In a 1.4 deg$^{2}$ sample
of the survey, approximately 87\% of the sources with 4 or more counts
have an optical counterpart to R$\sim26$ mag. As part of a larger
program of optical spectroscopy of the NDWFS Bo\"otes area, spectra
have been obtained for $\sim900$ of the X-ray sources, most of which
are QSOs or AGN. 
\end{abstract}


\altaffiltext{1}{Harvard-Smithsonian Center for Astrophysics, 60 Garden St., Cambridge MA 02138}

\altaffiltext{2}{National Optical Astronomy Observatory, 950 N. Cherry Ave.,Tucson Az 85719}

\altaffiltext{3}{Ohio University Department of Physics and Astronomy, Athens OH 45701}

\altaffiltext{4}{Steward Observatory, University of Arizona, 933 N. Cherry Ave., Tucson AZ 85751}

\altaffiltext{5}{The Ohio State University, Department of Astronomy, Columbus OH 43210}

\altaffiltext{6}{Department of Astrophysical Sciences, Princeton University, Peyton Hall, Ivy Lane, Princeton, NY 08544}

\section{$ $Introduction}

Many recent extensive multi-wavelength surveys of the properties of
galaxies and quasars probe only two cosmological regimes -- the very
local and the very distant. Local surveys (e.g., SDSS \citealt{2000AJ....120.1579Y}
or 2dF \citealt{2001MNRAS.328.1039C}) primarily select relatively
nearby galaxies and include cosmologically distant AGN, if they are
very luminous. Very deep surveys, like the Hubble Deep Fields, (HDF-N
\citealt{1996AJ....112.1335W}, HDF-S \citealt{2000AJ....120.2747C}),
the GOODS survey (\citealt{2004AdSpR..34..661C}), or the Chandra
Deep Fields (CDF-N \citealt{2001AJ....122.2810B}, CDF-S \citealt{2002ApJS..139..369G})
cover so little area (solid angle) that they primarily study distant
galaxies and extremely faint AGN. Medium depth multi-wavelength surveys
which cover the middle ground and allow us to explore the steady evolution
of galaxies and AGN with cosmic epoch exist (e.g., ChaMP, \citealt{2004ApJS..150...43G}),
but none cover the large contiguous areas necessary for detailed studies
of clustering and environment.

As a first step towards resolving this problem, the NOAO Deep Wide-Field
Survey (NDWFS, \citealt{1999prdh.conf..111J}) has obtained deep optical
($\mathrm{B_{w}}$, R, I) and near-infrared (K) images of two $\sim9$
square degree regions. In this paper we describe our X-ray imaging
survey of the Northern (Bo\"otes) field of the NDWFS. In addition
to the X-ray data, Bo\"otes has been imaged at radio (VLA FIRST,
\citealt{1996IAUS..175..499B} and WSRT \citealt{2002AJ....123.1784D}),
the mid-infrared (SST/IRAC, \citealt{2004ApJS..154...48E}), the far-infrared
(SST/MIPS, \citealt{2004AAS...204.4805S}), and the ultraviolet (GALEX,
\citealt{2003SPIE.4854..336M}). In addition to the imaging survey,
the AGN and Galaxy Evolution Survey (AGES, \citealt{Kochanek:2005})
has obtained redshifts for nearly 10,000 galaxies and quasars selected
from the NDWFS optical, X-ray, and infrared photometric samples.

We used the Chandra ACIS-I to survey 9.3 square degrees of the Bo\"otes
field by combining observations carried out as a collaboration of
Guest Observer (C. Jones PI) and Guaranteed Time Observer (S. Murray
PI) programs. A total of 126 contiguous exposures of $\sim5$ ksec
each were taken, allowing us to reach a limiting sensitivity of $\sim4\times10^{-15}\:\mathrm{erg\, cm^{-2}\, s^{-1}}$
in the energy range 0.5-7.0 keV . In this paper (Paper I) we outline
the data acquisition, reduction and point source detection procedures
(Section \ref{sec:X-ray-Observations-and}) leading to the XBo\"otes
catalogs denoted XB2 (sources with two or more counts) and XB4 (sources
with 4 or more counts). In Section \ref{sec:The-Distribution-of}
we discuss the general X-ray properties of the 4642 detected point
sources in XB2. The detailed X-ray source catalog will be presented
in Paper II (\citealt{Kenter:2004}), which also includes a list of
43 extended sources found in this survey. In particular, in Section
\ref{sub:The-Spatial-Distribution}, we note that the data are consistent
with no fluctuations in the source count density on the scale of the
ACIS-I field of view (16 arcminutes). In Section \ref{sub:Large-Scale-Structure}
we examine the angular correlation function of the X-ray sources,
and in Section \ref{sub:Optical-Matching-Against} we briefly discuss
the matches of X-ray sources with NDWFS optical sources. A full discussion
of the matches will be given in Paper III (\citealt{Brand:2004}).
The results of optical spectroscopy for X-ray selected sources are
discussed briefly in Section \ref{sub:Optical-Spectroscopy} and will
be the subject of future papers (e.g., \citealt{Kochanek:2005}).
One purpose of this early set of papers (I, II, and III) is to allow
us to make public the source catalogs in a timely fashion, while providing
an adequate reference for the methods used to generate them.

\section{X-ray Observations and Their Analysis \label{sec:X-ray-Observations-and}}

\begin{figure}[ht]
\begin{center}\includegraphics[%
  scale=0.5]{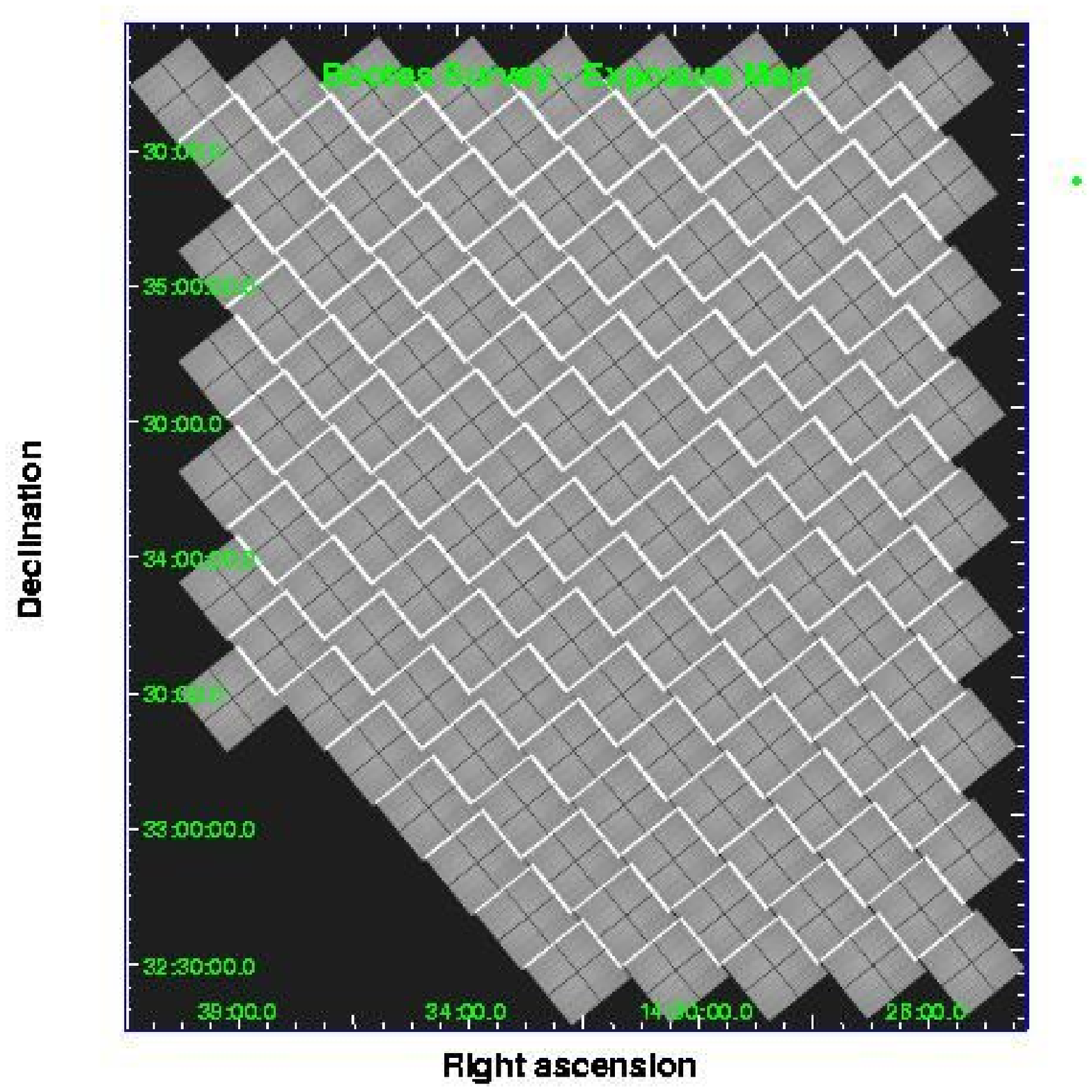}\includegraphics[%
  scale=0.33]{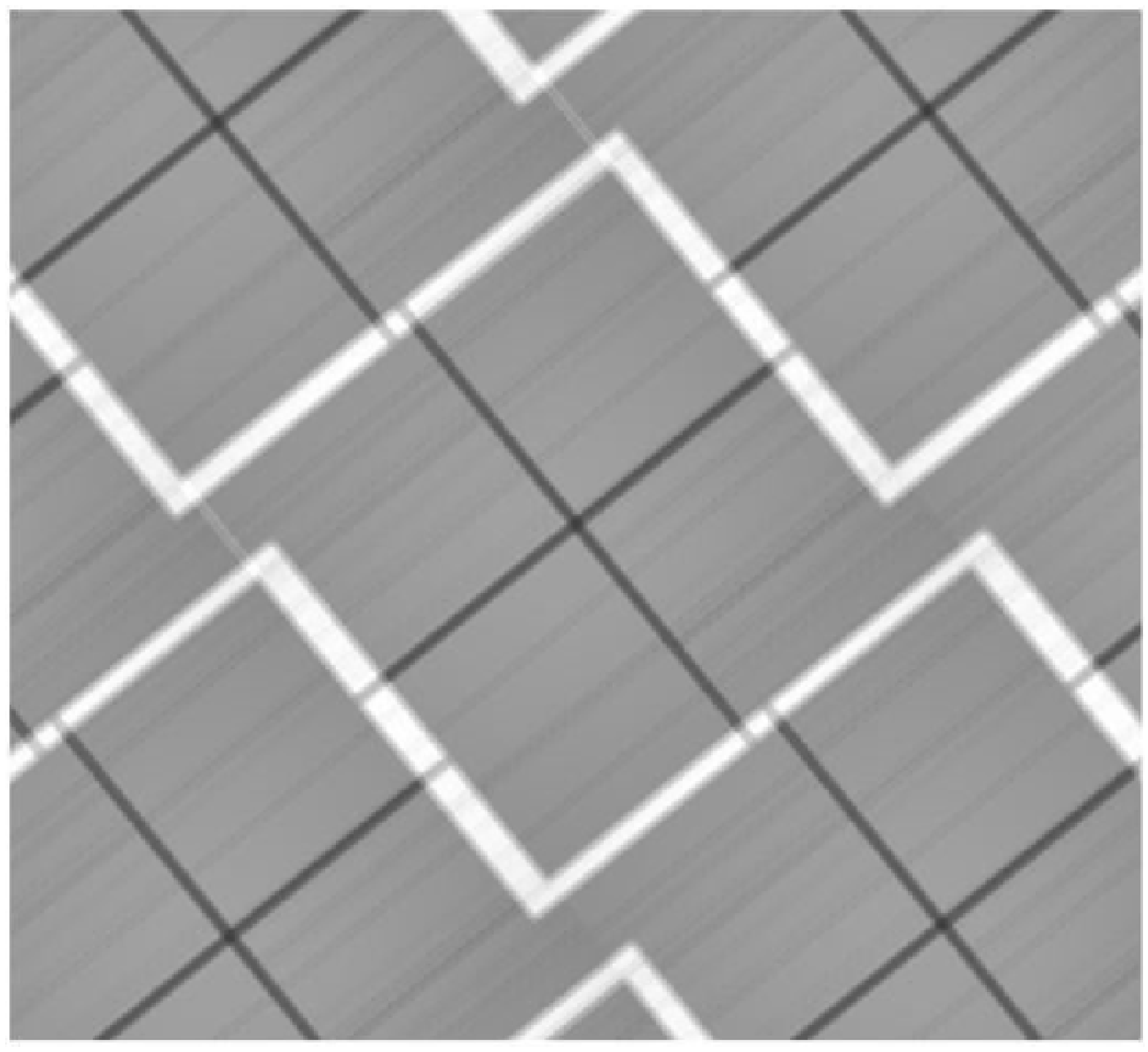}\end{center}

\caption{{\small Left: Exposure map for Bo\"otes Survey. There are 126 ACIS-I
fields, which overlap by less than 1 arc minute (the brighter regions).
Right: Expanded view near center. Due to telescope vignetting, the
center of each ACIS-I field has slightly (20\%) higher exposure than
the edge. The lower (darker) exposure region inside each ACIS-I field
is due to the gaps between the CCD chips. However, because the observatory
is dithered during an observation, the exposure in the CCD gaps is
not zero, but is about 50\% of the average exposure. \label{fig:exposure}}}
\end{figure}

The X-ray observations for this survey were carried out over a two
week time interval in March and April 2003, with the individual ACIS-I
field centers arranged so that the edges of fields overlapped by about
1 arc minute. The Bo\"otes field is centered at $\mathrm{RA}_{J2000}=14^{h}\,32^{m}$,
$\mathrm{Dec}_{J2000}=+35^{\circ}\,06^{'}$. The spacecraft roll angle
was maintained at a constant value for the entire set of 126 5 ksec
pointings, so that the resulting sky coverage would be as uniform
as possible. Figure \ref{fig:exposure} shows the exposure map for
the surveyand illustrates the overlap regions, as well as the effects
of telescope vignetting. The ACIS-I was operated in Very Faint Mode
to allow the best possible background rejection. In Table \ref{tab:obs}
we provide the observation details and Chandra observation identifiers.
These data are all publicly available through the Chandra X-ray Center
archive (Chandra Archive \url{http://asc.harvard.edu/cda}). A total
of 630 ksec of Chandra observing time was committed to this survey.

\begin{center}%
\begin{table}[ht]

\caption{Bo\"otes X-ray Observation Properties\label{tab:obs}\protect \\
}

\begin{center}\begin{tabular}{|c|c|c|c|}
\hline 
Chandra OBSIDs&
Type&
Date&
Locations\tabularnewline
\hline
\hline 
3596 - 3660&
GTO&
March 2003&
Northern\tabularnewline
\hline 
4218 - 4272&
GO&
April 2003&
Southern\tabularnewline
\hline 
4277 - 4282&
GO&
April 2003&
Southern\tabularnewline
\hline
\end{tabular}\end{center}
\end{table}
\end{center}

In Section \ref{sub:Data-Preparation} we discuss the reduction procedures
for the images, and in Section \ref{sub:Source-Detection} we discuss
the source detection procedures. Our procedures are very similar to
those used by Kenter and Murray (\citeyear{2003ApJ...584.1016K})
in their analysis of a smaller survey in the Lockman Hole area. Each
ACIS-I field was analyzed independently and the final source lists
are combined.

\subsection{Data Preparation \label{sub:Data-Preparation}}

The standard Chandra pipeline products are used with added processing
as follows. First the data are checked for periods of high background
or background flares following the CXC data preparation thread%
\footnote{{\small http://asc.harvard.edu/ciao/threads/filter}%
}. During these observations we had no major background flares, and
after screening, most of the individual observations have very much
the same exposure time. Almost all of the fields (110 out of 126)
have the same exposure time within 100 seconds, although the full
range is from 4250 to 5050 seconds. 

\begin{figure}[ht]
\begin{center}\includegraphics[%
  scale=0.75]{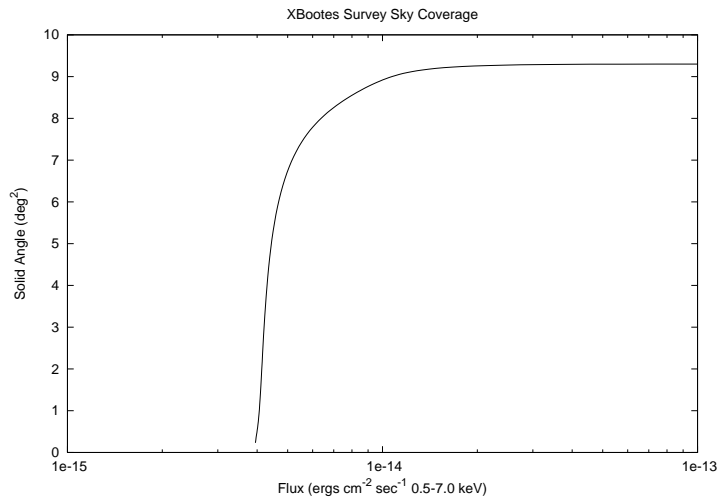}\end{center}

\caption{{\small Plot of the sky coverage for the XBo\"otes survey. The plot
gives the solid angle (in square degrees) as a function of the Total
band (0.5-7.0 keV) flux in $\mathrm{erg\, cm^{-2}sec^{-1}}$. The
nearly uniform exposure times of the 126 ACIS fields, accounts for
the rapid rise in coverage at the lowest fluxes. The roll over between
about 7 and 9 square degrees is due to the effects of vignetting and
the growth of the Chandra point spread function (PSF) at large off-axis
angles.\label{fig:coverage}}}
\end{figure}

The data are filtered using the very faint mode background algorithm
developed by Vikhlinin%
\footnote{{\small http://asc.harvard.edu/cal/Acis/Cal\_prods/vfbkgnd}%
}, which also removes afterglow events. Finally the data are filtered
by energy, limiting the energy range to 0.5-7 keV (Total band). The
final ACIS event files are then binned into image files, with a binning
factor of 4 (i.e., 4x4 ACIS pixels equal 1 image pixel 1.968 arc seconds
on a side). As noted above, these steps are performed on each survey
observation individually, resulting in 126 image files used for source
detection. In the Total band images, the typical background is $\sim1.1\times10^{-2}\,\mathrm{ct}/\mathrm{image\, pixel}$.
Of the 126 fields, there are 6 where the background is about a factor
of two higher than the rest. However, this higher background has a
negligible effect on point source detection efficiency within about
6 arc minutes of the field center, particularly for sources with $\geq4$
counts in the detection cell. Figure \ref{fig:coverage} shows the
cumulative sky coverage for the XBo\"otes survey calculated as in
\citet{2003ApJ...584.1016K}. Since the exposure times are all very
similar, the overall sky coverage rises rapidly as a function of flux
and exceeds 9 square degrees for a total band flux of $\geq10^{-14}\mathrm{erg\, cm^{-2}sec^{-1}}$. 

\subsection{Source Detection \label{sub:Source-Detection}}

We used the CIAO3.0.2 wavelet detect process (wavdetect) with the
{}``sigthresh'' parameter set to $5\times10^{-5}$ to detect point
sources in each of the data sets. This relatively high {}``sigthresh''
parameter value was set on the basis of simulations which showed that
for the short exposure times of our survey (and therefore low background)
the detection efficiency for faint sources is high and the spurious
detection rate is low (see Paper II, \citet{Kenter:2004} for details).
Because we analyze the fields independently, we may detect the same
source twice, if it lies in the overlap region (about 4\% of the total
area covered). We eliminate these duplications in our source list
by selecting the detection which has a smaller radial distance from
the aim-point (and therefore smaller point spread function). 

The results from running wavdetect (including the elimination of duplicated
sources and those with only one count) yields a combined list of 4642
point-like sources in the 0.5-7 keV energy band that have $\geq2$
counts within the 90\% encircled energy radius. We use this source
list as a starting point for further analysis to recalculate source
locations and fluxes from the original event data set. For each source
we extracted a 100x100 ACIS pixel {}``postage stamp'' at full ACIS
spatial resolution (0.492 arc seconds per pixel) centered on the wavdetect
position. We then recalculated the source position by iteratively
centroiding the events within the 90\% encircled energy region. We
used a circle to approximate the encircled energy region with radius
$R_{90}$ (in arc seconds) dependent on the square of the off-axis
angle of the initial source position%
\footnote{{\small The encircled energy radius (in arc seconds) is given by an
approximation to the relationship shown in the Chandra Proposer's
Observatory Guide (ACIS-I, for E=1.49 keV), using the functional form
$\mathrm{R}_{50}=0.423+0.0594\,\theta^{2}$ and $\mathrm{R}_{90}=0.881+0.107\,\theta^{2}$,
where $\theta$ is the off-axis angle in arc minutes.}%
}. 

Once the source location was determined, we counted the number of
events ($\mathrm{F_{50},\: F_{90}})$ within the 50\% ($R_{50}$)
and 90\% ($R_{90}$) encircled energy regions. If the source had at
least $\mathrm{F}_{50}\geq5$ counts inside $\mathrm{R}_{50}$, we
estimated the uncertainty in the source location as $\mathrm{C_{err}=R_{50}/(\sqrt{F_{50}}-1)}$,
i.e., the 50\% encircled energy radius divided by the approximate
counting uncertainty. For sources with $F_{50}<5$ counts, we set
the centroid uncertainty to be $\mathrm{C_{err}=R_{50}}.$ Finally,
for all sources, we set a floor to the centroid uncertainty of $\mathrm{\mathrm{C}_{err}}=1.5$
arc seconds to account for systematic errors not included in the above
analysis. These systematic effects include the non-circular shape
of the Chandra/ACIS point spread function (PSF) for sources that are
off-axis, the approximations associated with estimating the PSF size,
and residual astrometric errors in transferring detector coordinates
to the sky%
\footnote{{\small http://asc.harvard.edu/ca/ASPECT/celmon}%
}. Using the source positions obtained from the Total band (0.5-7 keV)
analysis, we also compute the total number of counts within the 50\%
and 90\% encircled energy regions in the Soft band (0.5-2 keV) and
the Hard band (2-7 keV), and from these generate the source hardness
ratio ($HR=(F90_{hard}-F90_{soft})/(F90_{hard}+F90_{soft})$). Using
the exposure map data from the pipeline processing, we also compute
the effective exposure time for each source, as well as the effective
area fraction relative to being on-axis (this provides the correction
for vignetting). The centroid uncertainties turn out to be fairly
conservative, crudely corresponding to about a 90\% confidence interval.
We will be able to quantify this more accurately as we accumulate
more spectroscopic identifications. It is already clear from the optical
matches (see \citealp{Brand:2004}) that the 1.5 arc second floor
to $\mathrm{C}_{err}$ (especially for sources close to the optical
axis) is generous.

\section{The Distribution of Point Sources\label{sec:The-Distribution-of}}

The source catalog, giving source positions and X-ray properties,
is presented in Paper II (\citealp{Kenter:2004}). We note here that
there were 4767 candidate sources from the CIAO wavdetect analysis.
From the processing described in the previous section, we find a total
of 4642 sources with greater than or equal to 2 counts in the 90\%
encircled energy region in the 0.5-7 keV Total energy band. We call
this the XB2 catalog. A subset of this catalog with sources of 4 or
more counts is denoted as the XB4 catalog and contains 3293 sources.
As a result of simulations, we estimate that the XB2 catalog contains
$\sim14$\% spurious sources, whereas the XB4 catalog contains less
than 1\% spurious sources. Here we discuss the spatial distribution
of the sources (i.e., their projected sky distribution), along with
a brief introduction to the optical matches and subsequent redshift
survey.

The 125 candidate sources with fewer than 2 counts within $R_{90}$,
which are not included in XB2, were examined individually and most
(114 of 125) were found to have only one event within the 90\% encircled
energy region appropriate to the off-axis position. Since the wavdetect
threshold was set high ($5\times10^{-5}$), it is not surprising to
have false candidate sources at the level of about 1 per ACIS-I field.

\subsection{The Spatial Distribution of Sources\label{sub:The-Spatial-Distribution}}

\begin{figure}[ht]
\begin{center}\includegraphics[%
  scale=0.5]{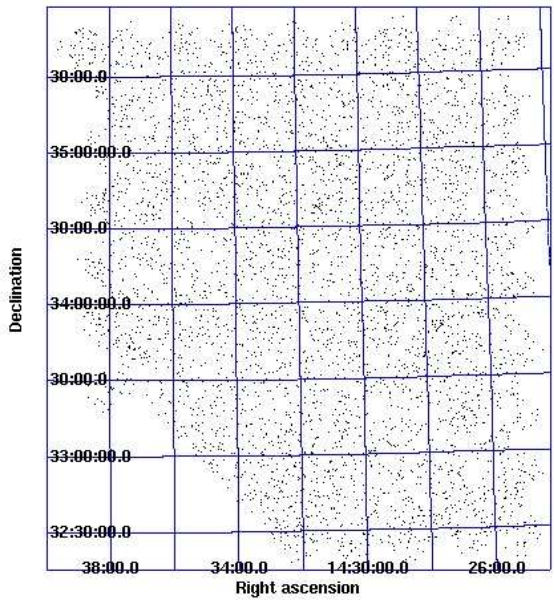}\end{center}

\caption{{\small Locations of the 4642 sources detected in the Bo\"otes Survey
with 2 or more counts within the 90\% encircled energy region of the
PSF.\label{fig:sources}}}
\end{figure}

There have been claims in deep surveys covering much smaller areas
for the detection of large scale structure in the distribution of
sources (e.g., \citealt{2003ApJ...585L..85Y}, \citealt{2002ApJS..139..369G}).
In contrast, the ChaMP found source distributions consistent with
Poisson statistics in 62 disjoint Chandra fields (\citealt{2004ApJ...600...59K}).
While our survey is not as deep, it is wide, contiguous in area, spatially
uniform, performed with a single instrument setup, and contains many
more sources that can be divided into spatial regions. Figure \ref{fig:sources}
shows the spatial distribution of the full set of Total band (0.5
- 7 keV) sources (XB2) on the sky, where each source is represented
by a dot (regardless of flux). No obvious significant spatial structure
is present. %

\begin{figure}[ht]
\begin{center}\includegraphics[%
  scale=0.8]{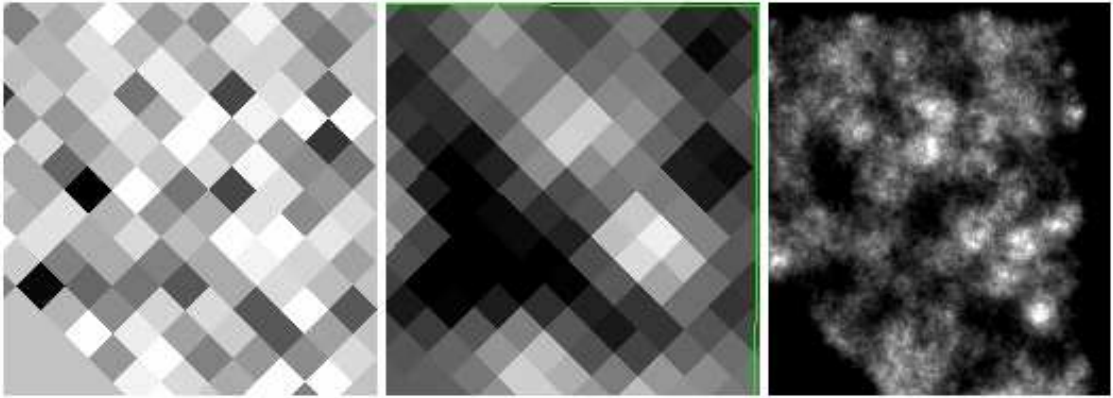}\end{center}

\caption{{\small Spatial distribution of sources per ACIS field where lighter
color means higher density. Left panel shows the {}``raw'' data,
i.e., the number of sources per ACIS-I field. The middle panel shows
the same data slightly smoothed (with a Gaussian using a sigma of
one ACIS FOV). The right panel is the image from Figure \ref{fig:sources}
adaptively smoothed with a 38 count threshold (the mean number of
sources per field). The structures seen in the middle panel are qualitatively
the same in this smoothed image.\label{fig:spatial}}}
\end{figure}

\begin{figure}[ht]
\begin{center}\includegraphics[%
  scale=0.5]{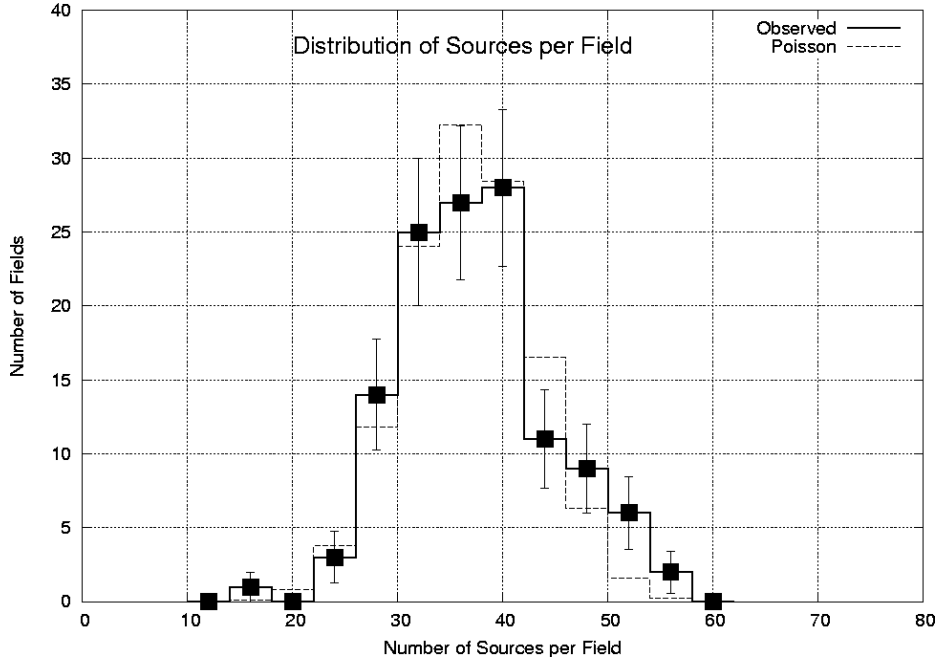}\end{center}

\caption{{\small Distribution of sources per field in XBo\"otes Survey. The
solid curve shows the number of fields in XBo\"otes as a function
of the number of sources (0.5 - 7 keV band, $\geq2$ counts) detected
in that field. The dashed curve is the expected Poisson distribution
for the mean number of sources per field 36.84) and the number of
fields (126).\label{fig:poisson}}}
\end{figure}

Figure \ref{fig:spatial} shows three different approaches to spatially
smoothing the distribution of sources. On the left is a {}``raw image''
corresponding to the number of sources in each ACIS-I field (pixels
are $\sim16$ arcminute on a side), where the color scale is darkest
for the lowest number of sources. The center image is the result of
Gaussian smoothing the raw image with a sigma corresponding to one
ACIS-I field. There appear to be spatial correlations (i.e., large
scale structures) where the source density is below the average (darker
areas that run NE to SW) and above the average (brighter areas to
the N and W of the dark patches). On the right we show an adaptively
smoothed image from Figure \ref{fig:sources}, where the top-hat smoothing
filter has an adaptive size needed to accumulate 38 counts (the average
number of sources per ACIS-I field). Structures similar to those seen
in the middle panel are also evident in this image. However, the significance
of these structures, estimated from the net excess (or deficit) of
sources compared with the average over the structure sizes, is only
2 to 2.5 sigma. The structures seen in Figure \ref{fig:spatial} are
typical in appearance and magnitude to structures observed in simulations
with random distributions of the same numbers of sources. We can quantify
this result by examining the counts-in-cells of the 126 ACIS-I fields
of view compared to the Poisson distribution for a mean of 36.84 sources
per field (4642/126). As shown in Figure \ref{fig:poisson} there
is no significant difference between the observed and the Poisson
distributions. A $\chi^{2}$ test gives a reduced value of 0.90 for
12 degrees of freedom. 

While it would be expected that the X-ray selected AGN are good tracers
of the cosmic web and large scale structure, it is not very surprising
that we find no evidence for large scale structure based on this technique.
Even with $\sim4500$ sources, it would take quite large amplitude
structures to overcome the statistical fluctuations after dividing
the data into 126 spatial bins. At $z\sim1$, an ACIS-I field corresponds
to about $8h^{-1}$ Mpc, a scale on which the three-dimensional correlation
function should have an amplitude near unity \citep{2002MNRAS.335..459C}.
In projection, however, we have averaged over the line-of-sight distance
which is 450 times larger than that scale size, and hence we would
expect fluctuations of only order $1/\sqrt{450}\sim5\%$, which are
too small to see against Poisson noise. Clear detections of large
scale structure on these scales will require redshifts (either spectroscopic
or photometric), so that the correlations will not be washed out by
projection effects along the large line-of-sight. In section \ref{sub:Optical-Spectroscopy}
we discuss some preliminary results on the redshift distribution of
a sample from the XB4 catalog that were obtained from the AGN and
Galaxy Evolution Survey (AGES, \citealt{Kochanek:2005}), which (when
complete) will yield accurate spectroscopic redshifts for about half
of the sources in the XB4 catalog as well as some of the optically
bright sources at lower flux from the XB2 catalog.

\subsection{Large Scale Structure and Angular Correlation\label{sub:Large-Scale-Structure}}

Statistical evidence of large scale structure (LSS) in X-ray surveys
has been reported by \citet{1995ApJ...455L.109V}, \citet{2001ApJ...551..624G},
\citet{2003ApJ...585L..85Y} and others. Previous X-ray surveys looking
for LSS typically have been limited to narrow field or disjoint serendipitous
surveys with spatially varying levels of sensitivity. The ROSAT survey
has further been subject to systematic errors and biases, due to poor
angular resolution (\citealt{1995ApJ...455L.109V}.) These Chandra
XBo\"otes data are unique in that they constitute the widest (9.3
deg$^{2}$), contiguous X-ray field yet observed to such a faint flux
limit as $4\times1\mathrm{0^{-15}erg\, cm^{-2}\, s^{-1}}$. The angular
resolution, uniform coverage and large contiguous field of view allow
us to search for evidence of structure on both small and large angular
scales, relatively free of edge effects and position inaccuracy biases.

In the absence of redshift information, we used the Landy and Szalay
(1993) estimator to determine the angular correlation function of
the X-ray sources. The random catalogs required for the estimate were
generated by drawing sources from the CDS logN-logS distribution (\citealt{2001ApJ...551..624G}),
and using the MARX (\citealt{1997adass...6..477W}) simulator for
the response of the Chandra X-ray Observatory. MARX simulates telescope,
detector and observatory features such as quantum efficiency, inter-CCD
chip gaps, vignetting, PSF and spacecraft dither. We have verified
that our Monte Carlo simulations reproduce many features present in
the real data, such as the fall off in detection sensitivity with
off axis angle.

\begin{figure}[ht]
\begin{center}\includegraphics{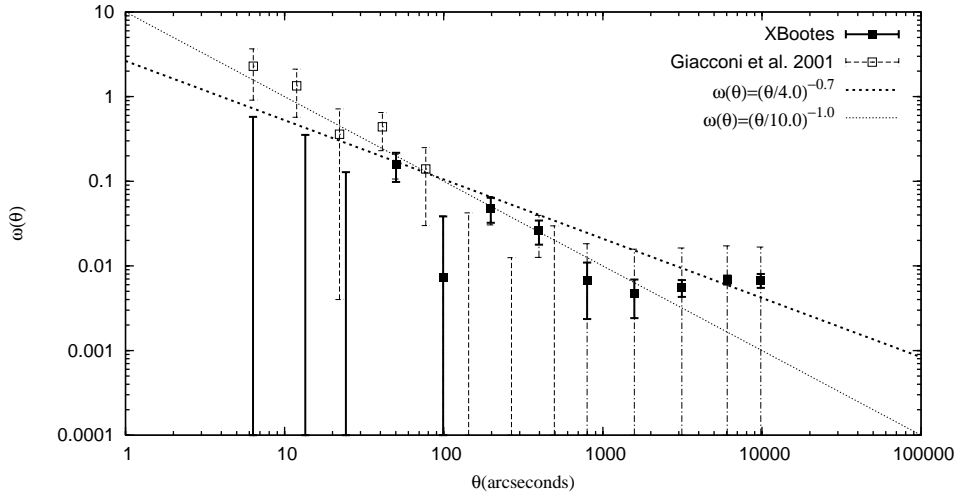}\end{center}

\caption{{\small The angular correlation function $\omega(\theta)$ for the
XBo\"otes Survey (filled squares and solid lines) derived using the
Landy and Szalay statistic $\omega(\theta)=(DD-2DR+RR)/RR$ and correcting
for the integral constraint. We find a positive angular correlation
on scales of $\geq1$ arcminute. These results are consistent with
the power law curve (shown as the heavy dotted line with index -0.7
and $\theta_{0}=4$ arcseconds) of \citet{1995ApJ...455L.109V} and
a similar power law (shown as the light dotted line with index -1.0
and $\theta_{0}=10$ arcseconds) of \citet{2001ApJ...551..624G}.
Also plotted, as the open squares and dashed lines, are the results
from \citet{2001ApJ...551..624G} (their Figure 6), which are consistent
with our results and the power law curves shown. \label{fig:ang_corr}}}
\end{figure}

In total, we simulated the entire XBo\"otes field sixteen times,
detecting $42,000$ simulated sources. The simulations reflect all
biases and features of the real XB2 data set. Taking into account
the integral constraint (\citealt{1977ApJ...217..385G}, \citealt{1991ApJ...380L..47E}),
we find a positive angular correlation in the full XB2 source catalog
on scales of $\geq1$ arcminute, as shown in Figure \ref{fig:ang_corr}.
The error bars shown are based on Poisson statistics ($\Delta\omega=[\omega(\theta)+1]/\sqrt{DD}$
c.f.,  \citet{2003psa..book.....W}). %
\footnote{Other estimators of errors and the effects of correlation are discussed
in the literature, see \citet{1996ApJ...473....7C} for a comparison
of these estimates. If these correlation effects are correct, then
the errors at the larger separation angles could be as much as 5 times
greater, while at the smaller separation angles the errors might be
up to 2 times larger. We indicate estimates for correlation effects
by the dot-dashed lines that extend the Poisson error bars in Figure
\ref{fig:ang_corr}%
} The results for $\theta\geq1$ arc-minute are consistent with those
previously given by \citet{1995ApJ...455L.109V}, who reported a correlation
described by the power law, $\omega(\theta)=(\frac{\theta}{\theta_{0}})^{-0.7}$
with $\theta_{0}\sim4$ arcseconds. On smaller scales the number of
source pairs in our data set is small and we have only upper limits.
However, we have plotted the correlation results from \citet{2001ApJ...551..624G}
(their Figure 6) which extend to small angles and are consistent with
our results. Also plotted is the somewhat steeper power law that \citet{2001ApJ...551..624G}
show in their figure. Our results are fully consistent with these
previous correlations.

\subsection{Optical Counterparts from the NDWFS\label{sub:Optical-Matching-Against}}

\begin{figure}[ht]
\begin{center}\includegraphics[%
  scale=0.5]{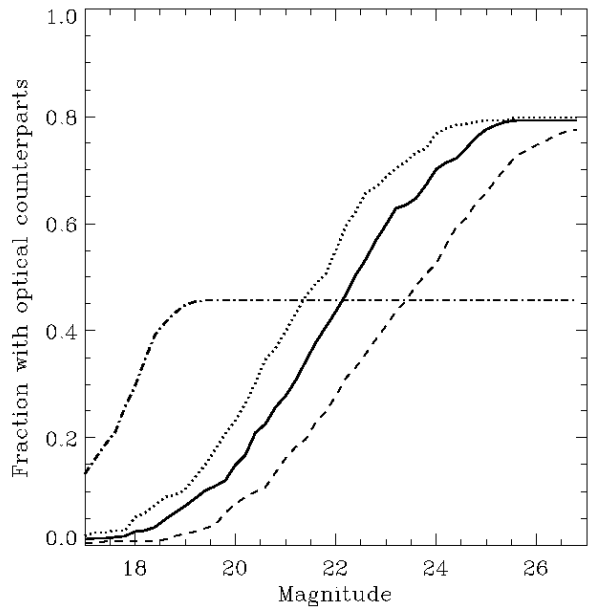}\end{center}

\caption{T{\small he cumulative fraction of X-ray sources with optical counterparts
brighter than a given magnitude (Vega) in the $R$-band (solid line),
$B_{W}$-band (dashed line), $I$-band (dotted line) and $K$-band
(dot-dashed line), taken from \citet{Brand:2004}\label{fig:fraction}}}
\end{figure}

The XBo\"otes X-ray survey field was chosen the cover the same region
as the Bo\"otes field of the NOAO Deep Wide-Field Survey: a deep
optical and near-IR imaging survey designed to study the formation
and evolution of large scale structure (\citealt{1999prdh.conf..111J}).
In this section, we present preliminary results on the identification
of the optical $(B_{W},\, R,\, I)$ and near-IR $(K)$ counterparts
to the XB4 catalog in a $1.4\,\mathrm{deg^{2}}$ sub-region of the
full Bo\"otes area. There are 481 X-ray sources in the sub-region
of which we expect less than 1\% to be spurious (\citealt{Kenter:2004}).
We find an optical counterpart for 87\% of the XB4 X-ray sources.
We assign an X-ray source to have an optical counterpart if there
is at least one optical source with a detection in at least one optical
band in the region enclosed by the X-ray positional errors ($\mathrm{C_{err}}$;
Section \ref{sub:Source-Detection}). The NDWFS positions have small
astronomical uncertainties (typically $0.1-0.4$ arcseconds, RMS)
that can be considered negligible in comparison to the positional
error of the X-ray sources (Section \ref{sub:Source-Detection}).
Figure \ref{fig:fraction} shows the cumulative fraction of matches
brighter than a given magnitude in the $B_{W},\, R,\, I$ and $K$
bands. The $K$-band catalog is not as deep as the optical catalogs
and therefore does not match the depth of XBo\"otes, resulting in
a smaller fraction of matches. In \citet{Brand:2004}, we present
a more sophisticated Bayesian identification scheme which self-consistently
evaluates the probability of each of the optical sources surrounding
an X-ray source being more probable than a 'background' source. We
also present the matched X-ray / optical catalog for the entire survey
area. 

\subsection{$ $Optical Spectroscopy\label{sub:Optical-Spectroscopy}}

The AGN and Galaxy Evolution Survey (AGES, \citealt{Kochanek:2005})
targeted all of the XB4 sources matched to an optical source brighter
than R=21.5 mag. Fainter X-ray sources from the XB2 catalog were targeted
if there were otherwise unallocated fibers. Spectra were obtained
with the MMT using the 300-fiber Hectospec robotic spectrograph (\citealp{1998SPIE.3355..285F})
in Spring and Summer of 2004. We \citep{Kochanek:2005} obtained optical
spectra for 1231 X-ray selected targets, and the preliminary reduction
of these spectra have resulted in 892 well-determined redshifts and
preliminary spectral classifications for these sources. While more
work is yet to be done, we find that, as expected, the detection of
X-rays preferentially identifies AGN from the {}``sea'' of galaxies
in the NDWFS, and that at the sensitivity level reached in this survey,
most of the AGN are at a redshift of about 1. Of the 892 X-ray objects
with preliminary redshifts, 25 are stars, 249 are classified as galaxies,
43 are classified Sy$\,$1/2 galaxies, and 575 classified as QSO/Sy$\,$1
galaxies. These classifications are based on template matching of
the extracted spectra (about 6$\textrm{Å}$ FWHM resolution) using
either a galaxy template (\textit{i.e.}, absorption lines) or an emission
line template. The AGN/Sy 1 have broad lines and the Sy 1/2 have either
{[}NeV{]} 3426$\textrm{Å}$ at > 2.5$\sigma$, or {[}N II{]} > H$\alpha$,
{[}OI{]} 6300$\textrm{Å}$ exists and {[}O III{]} 5007$\textrm{Å}$
$>2\times\mathrm{H}\beta$. It is possible that some of the objects
classified as galaxies have lower significance AGN emission lines
which were not flagged by this analysis.

\begin{figure}[ht]
\begin{center}\includegraphics[%
  scale=0.30]{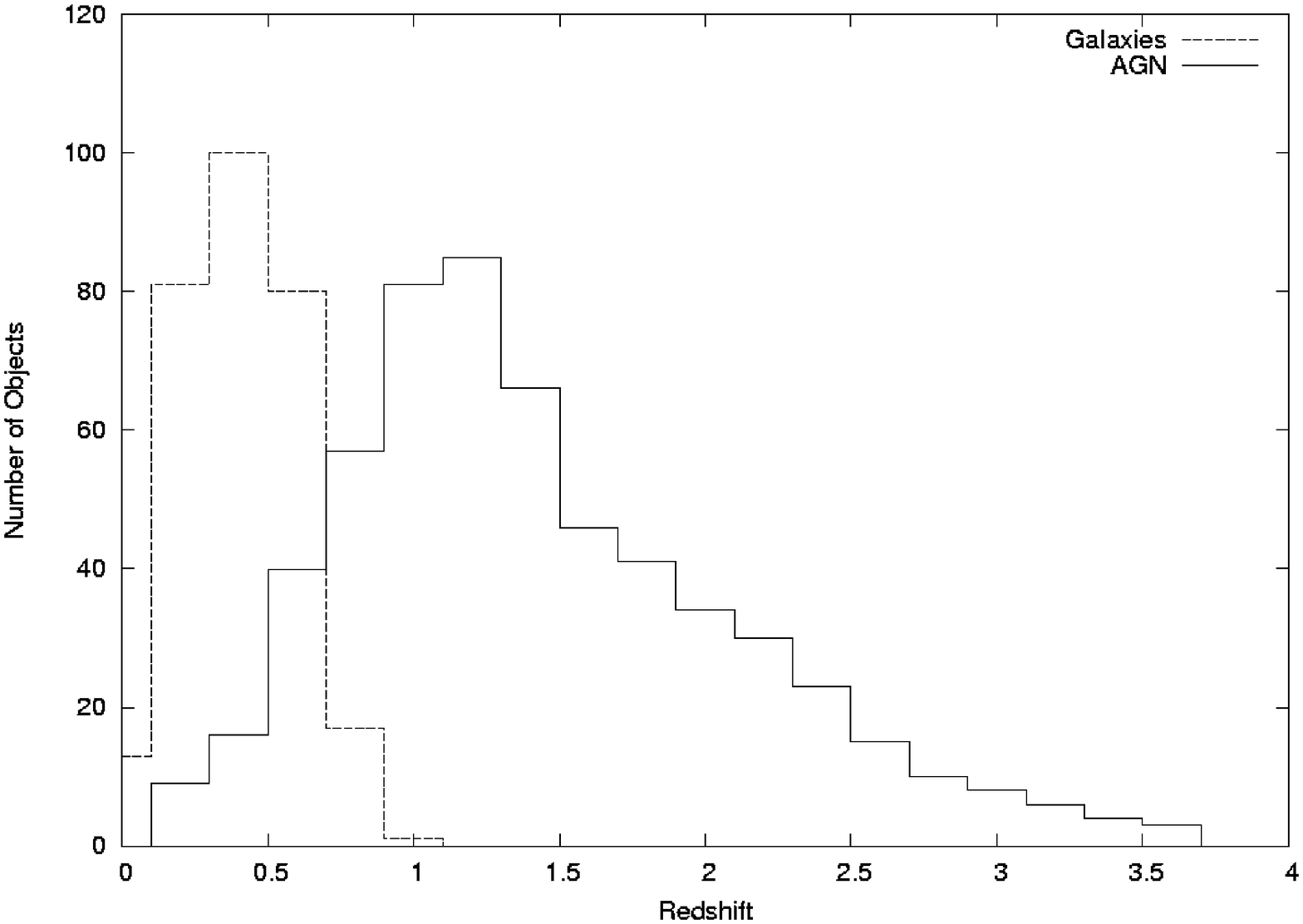}\includegraphics[%
  scale=0.30]{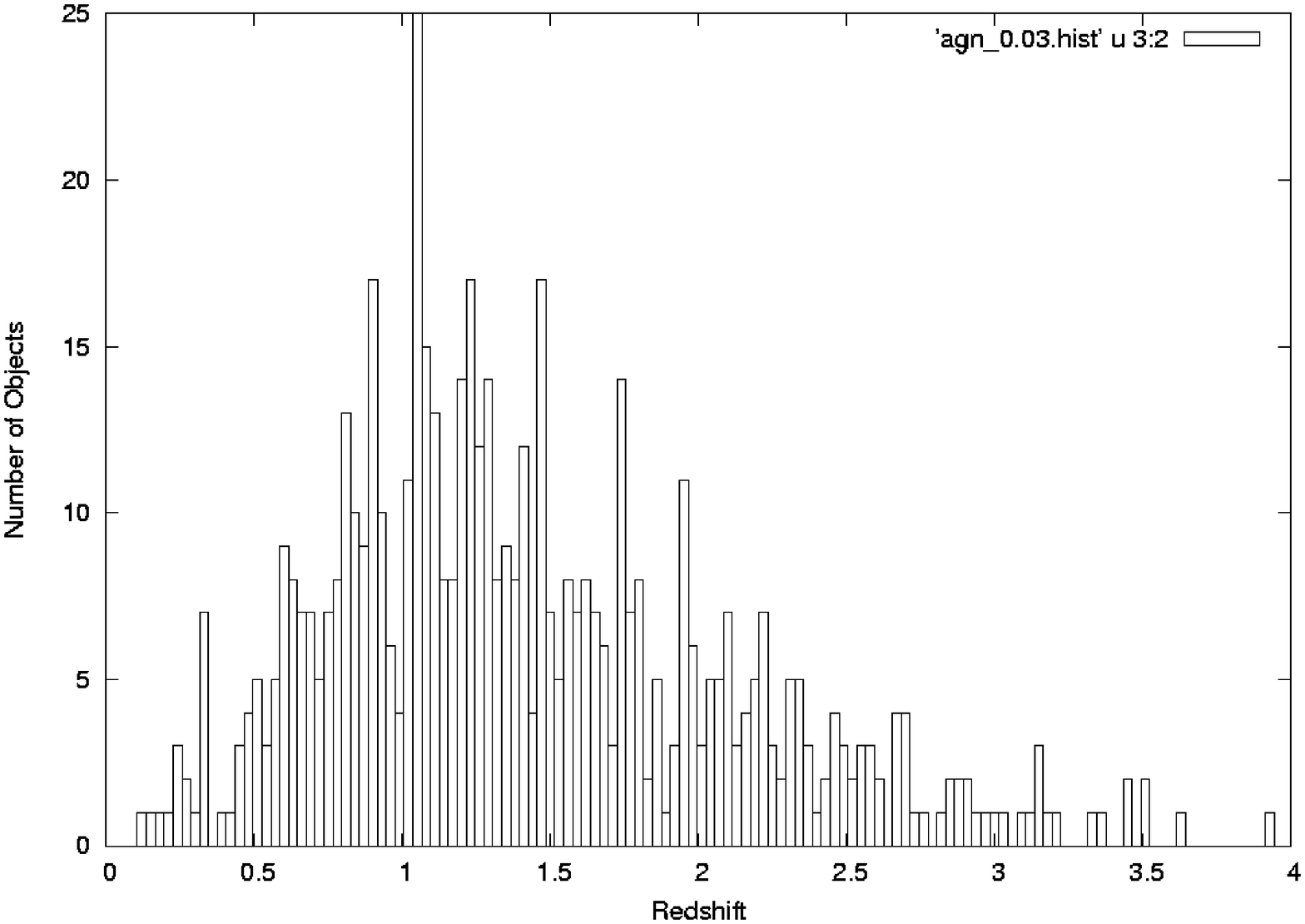}\end{center}

\caption{{\small Redshift distributions. Left: for galaxies (dashed) and AGN
(solid) from AGES measurements using the MMT/Hectospec.\label{fig:agn_z_dist}.
Right: More finely binned (0.03) for AGN only showing possible spikes
that might correspond to large scale structure features.}}
\end{figure}

The redshift distributions for galaxies (not flagged as any kind of
AGN), and AGN (QSO/Sy 1/Sy 2) are shown on the left panel of Figure
\ref{fig:agn_z_dist}. \citet{2004ApJ...616..123T} and
\citet{2005AJ....129..578B} plot redshift distributions for the AGN
found in the Chandra Deep Fields showing peaks near
z=1. \citet{2005AJ....129..578B} give the median redshift as a
function of X-ray flux, and for the flux range of the XBo\"otes
survey, these are all near z=1. The peak for the XBo\"otes AGN is also
peaked around z=1 consistent with these results.  With only a
preliminary separation of galaxy and AGN classes, it is not
appropriate to compare our redshift distribution in detail with those
for CDFS (e.g., \citealp{2004ApJS..155..271S}) or detailed models such
as those of \citet{2004ApJ...616..123T}. The long tail out to
redshifts as high as 3.9 is not surprising, as the survey area is
large enough to find rare very high luminosity AGN at large redshift.

The right panel of Figure \ref{fig:agn_z_dist} shows the distribution
of AGN with a finer binning (0.03 in z). There are several spikes
in the distribution which, if real, would indicate the presence of
large scale structure. We have used a one-sided statistical test of
Nulsen and Murray (2005, in preparaton) to search for significant
excesses in the redshift bins.%
\footnote{If the counts for a bin ($N_{2}$) are drawn from a population with
the same mean source density $\mu$ as its adjacent neighbors with
counts $N_{1}$ and $N_{3}$, then the likelihood of the excess is
measured by $A=P(\leq N_{1}\mid\mu)\, P(\geq N_{2}\mid\mu)\, P(\leq N_{3}\mid\mu)$.
To measure the significance of this (or any other) spike, we maximize
$A$ with respect to the source density $\mu$ (allowing for differences
in co-moving volume), to obtain $A=a$ ($a$ being the maximum value
for the joint likelihood $A$) for $\mu=\overline{\mu}$. If the counts
are drawn from populations with common mean density $\overline{\mu}$,
we can then calculate the cumulative probability, $P(A<a\mid\overline{\mu})$,
of observing this joint likelihood ($a$), or smaller. If this probability
is small, then the spike is \char`\"{}real\char`\"{}. %
} Given the spareness of the data available, we do not consider any
of the spikes in our current redshift distribution to be firm evidence
for large scale structure. However, once all of the X-ray selected
AGN have been observed as part of our ongoing AGES program, there
should be about twice the number of objects and therefore sufficient
statistics to identify redshift concentrations corresponding to actual
large scale structures.

\begin{figure}[ht]
\begin{center}\includegraphics[%
  scale=0.35]{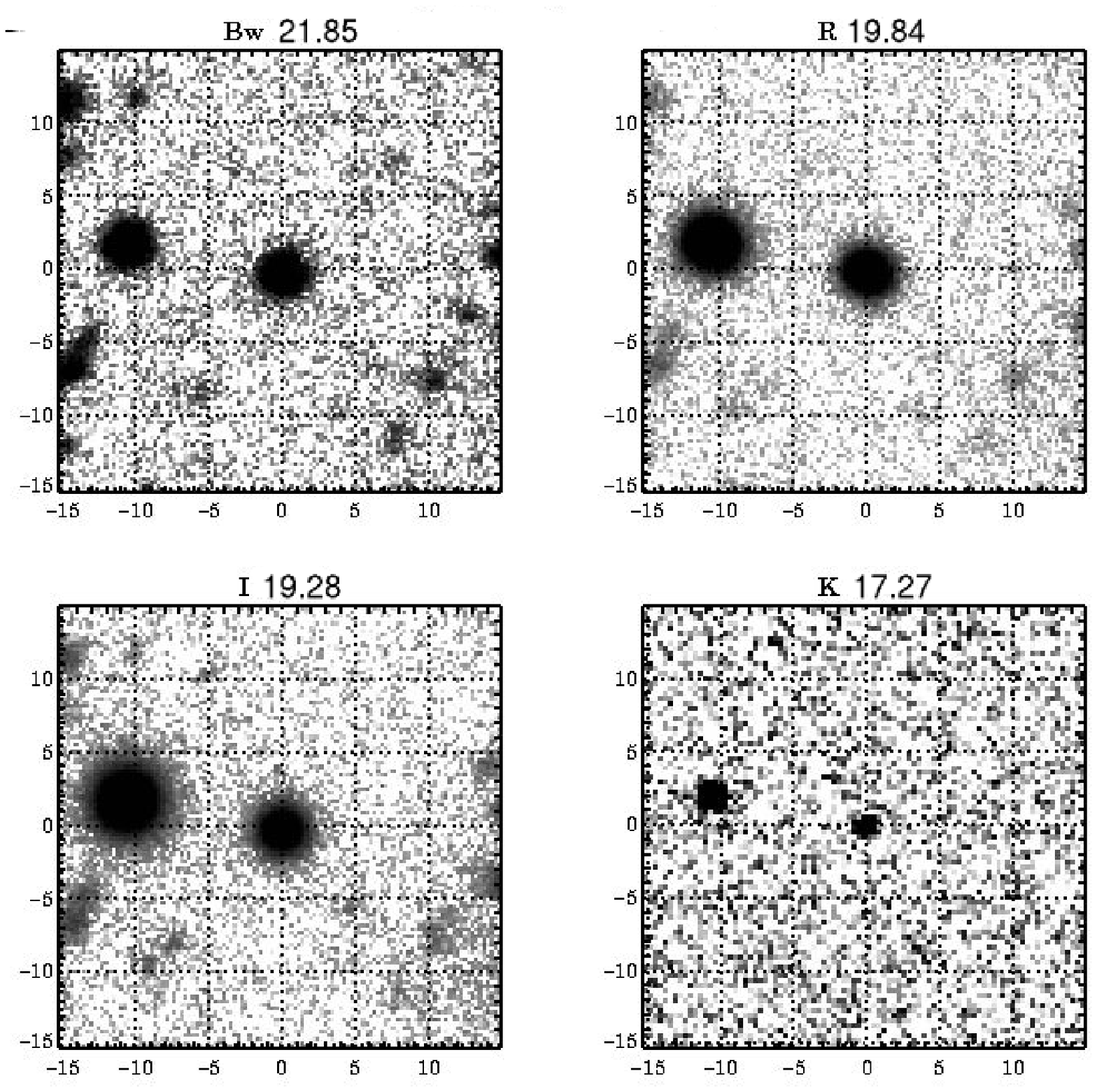}\includegraphics[%
  scale=0.35]{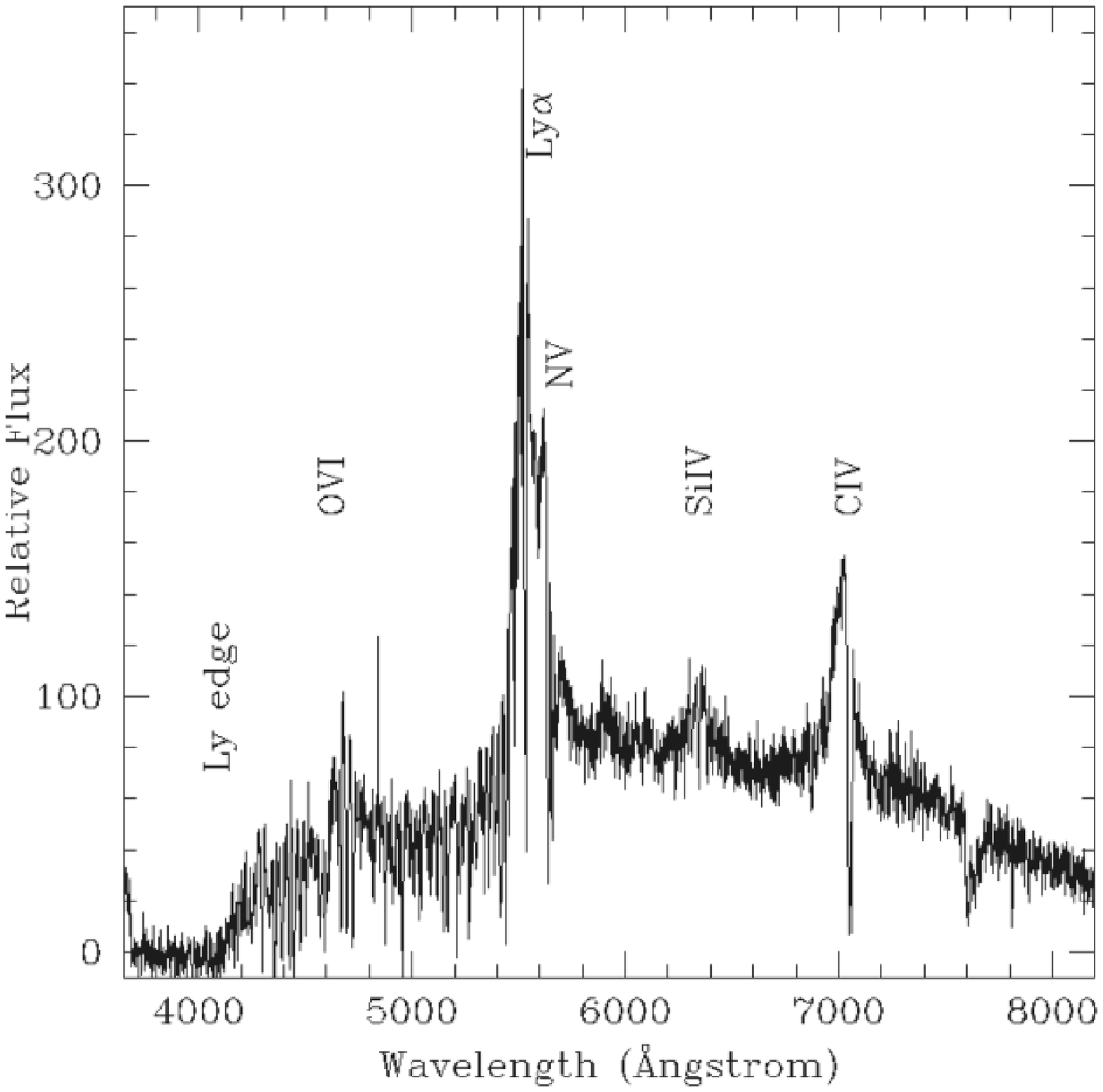}\end{center}

\caption{{\small Finding chart for NDWFS J142547.4+352719 (aka CXOXB J142547.4+352719),
a z=3.53 AGN and its preliminary spectrum.\label{fig:agn_spectra}}}
\end{figure}

The quality of the AGES spectra are illustrated in Figure \ref{fig:agn_spectra}
where we show the optical finding chart and preliminary spectrum for
one of the highest redshift sources, CXOXB J142547.4+352719 (aka NDWFS
J142547.4+352719), a quasar at z=3.53 which has 12 counts in the Total
band (0.5-7 keV), corresponding to a luminosity $\mathrm{L_{x}}=2.8\times10^{45}\mathrm{\, erg\, cm^{-2}s^{-1}}$
(using $\lambda\mathrm{CDM}$ with $H_{0}=71,\:\Omega_{m}=0.27,\:\mathrm{and}\:\Omega_{\lambda}=0.73$).

\section{Conclusions\label{sec:Conclusions}}

We have conducted a large area survey using 126 Chandra ACIS-I pointings
to cover 9.3 square degrees of the NDWFS in Bo\"otes. We find no
evidence for major deviations from a uniform sky density of sources
at the flux levels reached in $\sim5000$ seconds of Chandra observing
time, but there is some hint of spatial structure on a scale of several
Mpc which may be due to the X-ray selected AGN population tracing
out large scale structures. The two-point angular correlation for
$\theta\geq1$ arcminute does show the same power law correlation
as noted previously (e.g., \citealt{1995ApJ...455L.109V}, \citealt{2001ApJ...551..624G}). 

The X-ray survey and the NDWFS are well matched as evidenced by the
high (87\%) success rate (\citealt{Brand:2004}) of associating X-ray
sources with optical candidates. Follow up optical spectroscopy as
part of AGES (\citealt{Kochanek:2005}) has yielded good classification
and redshift results, with 892 preliminary redshifts out of 1231 targets,
most of which were brighter than R=21.5. There are hints in the binned
redshift distribution of these X-ray selected AGN for excesses at
several redshifts. However, the small number of objects per redshift
bin does not allow for these to be taken as firm evidence for large
scale structures. These initial redshifts, plus additional spectroscopic
redshifts expected from future AGES observations (and ultimately augmented
with a comparable number of photometric redshifts) will permit a statistically
interesting view of the spatial distribution of the X-ray selected
sources and their relationship to large scale structures as traced
by the galaxy spectroscopy.

\section{Acknowledgements}

This work was supported through the Smithsonian Institution and by
NASA Contracts NAS8-38248, NAS8-01130, NAS8-39073, NAS8-03060, and
NASA Grant GO3-4176A. This work was also supported by the National
Optical Astronomy Observatory which is operated by AURA, Inc, under
a cooperative agreement with the National Science Foundation. We appreciate
the excellent support we have received from the CXC Mission Planners
in carrying out these observations, the CXC Data Processing Team for
the pipeline data, the NDWFS Team for the optical observations and
data reduction, and the AGES Team in obtaining reduced spectra. We
would like to thank the anonymous referee for helpful suggestions
that improved this paper.


\end{document}